\documentclass[amsmath,amssymb,prl,aps,twocolumn,mathbbm,superscriptaddress]{revtex4}
\usepackage{amssymb}
\usepackage{graphicx}
\usepackage{times}
\usepackage{subfigure}

\begin{document}
\title{Quantum simulation of traversable wormhole spacetimes in a dc-SQUID array}
\author{Carlos Sab{\'\i}n}
\address{Instituto de F\'isica Fundamental, CSIC, Serrano 113-bis
  28006 Madrid, Spain}

\date{\today}

\begin{abstract}
We present an analog quantum simulator of spacetimes containing  traversable wormholes. A suitable spatial dependence in the external bias of a dc-SQUID array mimics the propagation of light in a 1D wormhole background. The impedance of the array places severe limitations on the type of spacetime that we are able to implement. However, we find that wormhole throat radius in the sub-mm range are achievable. We show how to modify this spacetime in order to allow the existence of closed timelike curves. The quantum fluctuations of the phase associated to the finite array impedance might be seen as an analogue of Hawking's chronology protection mechanism. 
\end{abstract}
\pacs{}
\maketitle

Quantum simulators \cite{reviewqs} are becoming increasingly popular as non-universal quantum computers with the potential of proving the long-sought {\it quantum supremacy} \cite{preskill}. In addition to this most practical application, quantum simulators have proven to be useful tools to explore the frontiers of quantum physics, ranging from open problems in well-established theories such as quantum field theory \cite{nonabelian} to untested physics whose observability is hard or dubious \cite{zitterbewegung,majoranas1,majoranas2,gws1,gws2} or even probably impossible \cite{tachions1, tachions2}.

Wormholes or Einstein-Rosen bridges are compelling mathematical objects appearing in some solutions of  Einstein's General Relativity equations. Since they provide a bridge between distant regions of spacetime, they have attracted a great deal of attention from a foundational viewpoint as well as at a pedagogical level \cite{einsteinrosen,wheeler, hartle, morristhorne, morristhorne2}. However, it seems that they do not appear {\it naturally} in our Universe and moreover there are reasons to expect that even a hypothetical manufacture must be forbidden \cite{hawking}. The stability of a wormhole relies on the use of exotic material violating the weak energy condition -namely, the existence of spacetime regions with negative energy density for some observers- and wormhole spacetimes may contain closed timelike curves (CTCs) \cite{morristhorne} which are typically deemed as incompatible with the physical principle of causality. However, at least at the quantum level it is possible to reconcile causality and CTCs \cite{deutsch}. Indeed, CTCs would boost the capabilities of quantum computers \cite{openctcs}- an observation that has motivated the interest of quantum simulation of CTCs \cite{timralph}.

\begin{figure}[h!]
\includegraphics[width=0.5\textwidth]{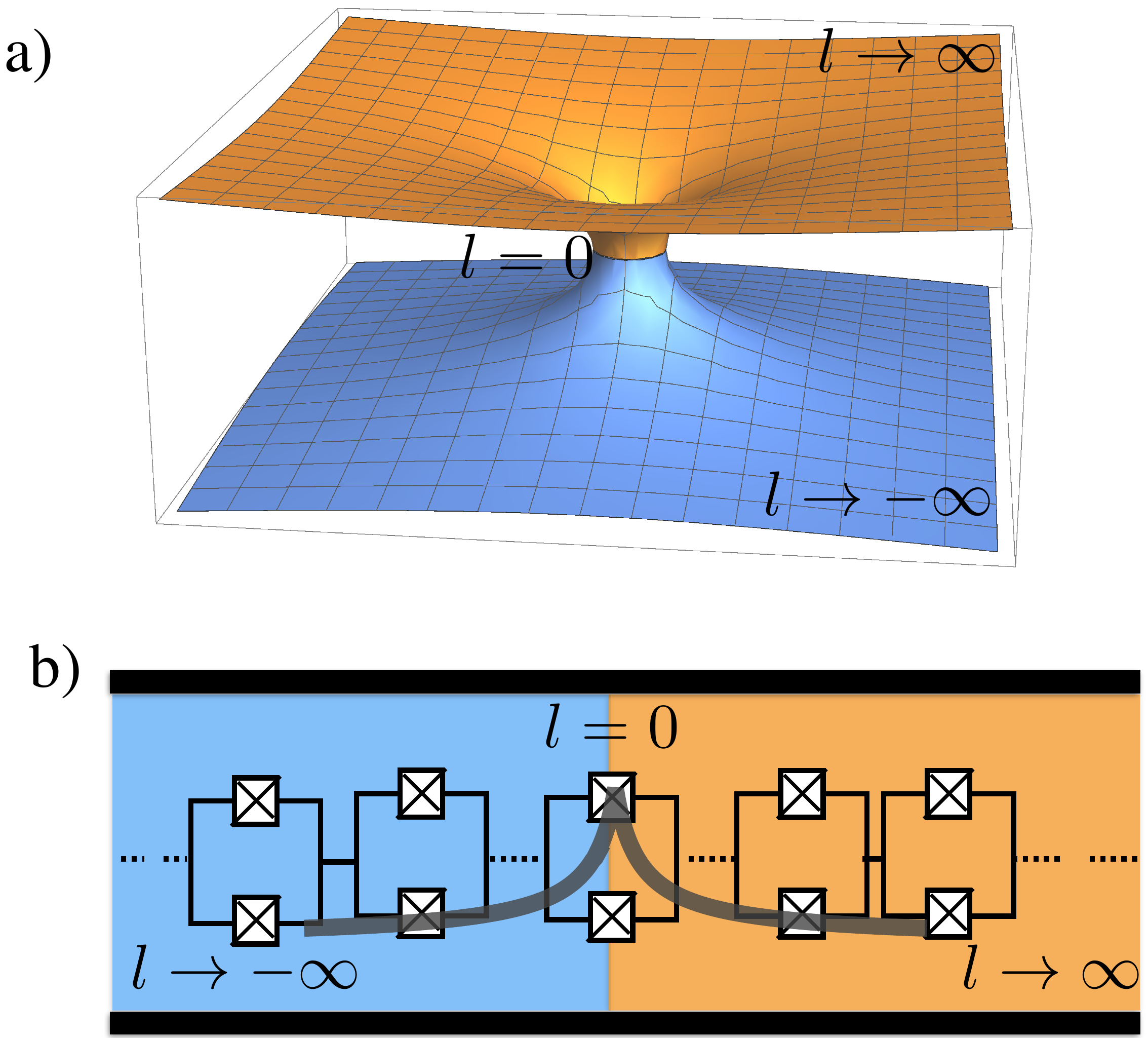}
\caption{(Color online) a) Embedding diagram of a traversable wormhole spacetime. Two asymptotically flat regions of spacetime at $l \rightarrow\pm\infty$ are connected by a throat centered at $l=0$, where $l$ is defined by the proper radial distance to the wormhole throat. b) An array of dc-SQUIDS embedded in a superconducting open transmission line. A suitable strongly inhomogenous external flux bias (gray) generates an effective speed of propagation for the electromagnetic quantum field, which can mimic the one in the spacetime depicted in a).}
\label{fig1}
\end{figure}

In this work we introduce a quantum simulator of traversable wormhole spacetimes by means of a supeconducting circuit architecture consisting of an array of dc-Superconducting Quantum Interferometric Devices (SQUIDs). Superconducting setups have already proven useful for the simulation of relativistic physics \cite{casimirwilson,casimirsorin,nationhawking,reviewjohansson}. We show that 
the wide tunability of the SQUIDs can be exploited to mimic the propagation of a microwave electromagnetic field near a one-dimensional (1D) wormhole, thus generating an effective wormhole spacetime for the quantum field. This is a remarkable difference with previous proposals \cite{hojaverde, rousseax} to simulate wormholes in metamaterials or in water, which are based on the {\it classical} Maxwell equations- a classical simulator of a wormhole for magnetostatic fields has been recently implemented in the laboratory with a magnetic metamaterial \cite{magneticwormhole}. We consider a paradigmatic family of traversable wormholes \cite{morristhorne2}, as well as a modification of them which can contain CTCs \cite{morristhorne}. We see that the electromagnetic impedance of the array places severe limitations on the simulated spacetime parameters, by generating quantum fluctuations of the superconducting phase in the surroundings of the simulated wormhole throat. We try to minimise the region of the array where the impedance is large, ideally to a single point representing the throat which in turn establishes a limit on the size of the simulated throat radius. These limitations coming from quantum phase fluctuations can be seen as an analogue of Hawking's chronology protection mechanism \cite{hawking}, where quantum effects prevent us from building spacetime geometries which might collide with the causality principle.   

Let us start by introducing the family of spacetime metrics considered in this work. A traversable wormhole spacetime can be characterised by \cite{morristhorne2}:
\begin{equation}
ds^2=-c^2\,e^{2\Phi(r)}dt^2+\frac{1}{1-\frac{b(r)}{r}}\,dr^2 +r^2(d\theta^2+\sin^2\theta d\phi^2), \label{eq:metric}
\end{equation}
where the {\it redshift function} $\Phi(r)$ and the {\it shape function} $b(r)$ are functions of the radius $r$ only. There is a value $b_0$ of $r$ at which $b\, (r=b_0)=r=b_0$, which determines the position of the wormhole's throat. Then, the {\it proper} radial distance to the throat is defined by \cite{morristhorne2} $l=\pm\int_{b_0}^r\,dr'(1-b(r')/r')^{-1/2}$, defining two different ¬Universes¬ or regions within the same Universe for $l >0$ (as $r$ goes from $\infty$ to $b_0$) and $l<0$ (as the non-monotonic $r$ goes back from $b_0$ to $\infty$). Thus, as $r\rightarrow\infty$ we have two asymptotically flat spacetime regions $l\rightarrow\pm\infty$ connected by the wormhole throat at $l=0$ ($r=b_0$). (See the embedding diagram in Fig. (\ref{fig1}a), which is obtained using standard embedding techniques \cite{morristhorne2}).

In this work, we will consider for simplicity that $\Phi(r)=0$ (massless wormhole). The properties of the wormhole will depend on the form of the shape function $b(r)$. In particular, as shown in \cite{morristhorne2} the parameters of this function can be adjusted in order to make traversability possible and convenient. We will consider some particular shape function later.

First, we will restrict ourselves to 1D spacetimes:
\begin{equation}
ds^2=-c^2\,dt^2+\frac{1}{1-\frac{b(r)}{r}}\,dr^2.\label{eq:metric2}
\end{equation}
In this way, we can exploit the invariance of the 1D Klein-Gordon equation under conformal factors \cite{birrelldavies}. This means that the dynamics of a 1D electromagnetic field in the spacetime given by Eq. (\ref{eq:metric2})  is totally equivalent to the one in  the following spacetime
\begin{equation}\label{eq:metric3}
ds^2=-c^2\,(1-\frac{b(r)}{r})\,dt^2+\,dr^2,
\end{equation}
since the line element in Eq. (\ref{eq:metric2}) differs to the one in Eq. (\ref{eq:metric3}) by the conformal factor $1/(1-b(r)/r)$ only. 

The spacetime given by the line element in Eq. (\ref{eq:metric3}) is a spacetime in which the speed of propagation of the electromagnetic field depends on the radius $r$ according to:
\begin{equation}\label{eq:light}
c^2(r)=c^2\,(1-\frac{b(r)}{r}),
\end{equation} 
which suggests that any experimental setup in which the effective speed of light of Eq. (\ref{eq:light}) can be produced is a suitable analog quantum simulator of a 1D wormhole spacetime. Note that Eq. (\ref{eq:light}) predicts that $c$ is exactly 0 at the throat, but larger than 0 at both sides of the throat. Thus light would experience an acceleration as traversing the throat. 

In this work we consider a dc-SQUID array embedded in an open transmission line \cite{casimirsorin,array, array2,array3}. The speed of propagation along the transmission line is given by $c=1/\sqrt{L\,C}$, where $C$ and $L$ are the capacitance and inductance per unit length respectively. We will assume that the number of embedded SQUIDs is large enough to consider that $C$ and $L$ are given by the capacitance and inductance of a single SQUID $C_s$ and $L_s$. If the SQUIDs area is small enough we can neglect their self-inductance. In this case and considering that the two Josephson junctions (JJ) of each dc-SQUID possess identical critical current $I_c$, we can treat any SQUID as a single JJ with a tunable inductance for frequencies well below the plasma frequency of the SQUID \cite{simoenthesis}: 
\begin{equation}\label{eq:squidind}
L_s(\phi_{ext})=\frac{\phi_0}{4\pi\, I_c\cos\frac{\pi\phi_{ext}}{\phi_0}\cos\psi}, 
\end{equation}
where $\phi_0=h/(2\, e)$ is the flux quantum, $\phi_{ext}$ is the external magnetic flux threading the SQUID and $\psi$ is the SQUID phase, which gives rise to a nonlinearity. We will remain within the linear regime, that is we can assume the approximation $\cos\psi\simeq1$ (we will comment on this in more detail below) and then the speed of light becomes
\begin{equation}\label{eq:squidspeed}
c^2(\phi_{ext})=c^2\cos\frac{\pi\phi_{ext}}{\phi_0}
\end{equation}
where we are denoting $c$ as the speed of light in the absence of external flux $c^2=c^2(\phi_{ext}=0)=1/(L_s(\phi_{ext=0})C_s)$. 

By inspection of Eqs. (\ref{eq:light}) and (\ref{eq:squidspeed}), we find that  a wormhole spacetime can be realised as long as the external magnetic flux has the following dependence on some variable $r$:
\begin{equation}\label{eq:externalflux}
\phi_{ext} (r)=\frac{\phi_0}{\pi}\operatorname{arccos} (1-\frac{b(r)}{r}).
\end{equation}

We will relate $r$ with an actual position coordinate in the laboratory in a particular example later. Before specialising to particular shape functions $b(r)$ it is important to make an important remark on Eq. (\ref{eq:externalflux}). There will always be a point at which $b(r)=r$ and then $\phi_{ext}=\phi_0/2$, determining the simulated wormhole's throat as a point of infinite inductance in the array. It is well-known that if all the SQUIDs of the array were biased with such a value of the external flux, the approximation $\cos\psi\simeq1$ would be no longer valid. Indeed, the array would not be in the superconducting state \cite{array}, since quantum fluctuations of the phase would become dominant, triggering a quantum phase transition to an insulating state. This is because in an SQUID array, the impedance $Z$ of a SQUID does not depend on the impedance $Z_E$ of the electromagnetic environment only (typically negligible) but also on the impedance of the rest of the array $Z_A$, which depends on the external flux. In some experiments, the parameters of the array are chosen precisely to use the array as a high impedance electromagnetic environment \cite{array, array2, array3}. We pursue the opposite goal here. More specifically, we need to make sure that $Z_A/R_Q\leq1$ everywhere, where $R_Q=h/(4e^2)$ is the resistance quantum. In our case, the impedance is $r$-dependent and will be given by \cite{array,nationhawking}: 
\begin{equation}\label{eq:impedance}
\frac{Z_A(r)}{R_Q}=\sqrt{\frac{2\pi\,e^2}{\phi_0C_0I_c\cos\frac{\pi\phi_{ext} (r)}{\phi_0}}},
\end{equation}
where $C_0$ is the capacitance-to-the-ground of the transmission line. Assuming for instance the realistic values $I_c=10\,\operatorname{\mu A}$, $C_0=0.1 \operatorname{pF}$, we find that $Z_A/R_Q\simeq 1$ for $\phi_{ext}\simeq0.45\phi_0$. Therefore, we would like to minimise the region of the array where the flux takes values above this threshold. Ideally, we would like to have a single SQUID only above the threshold - the one defining the throat. If that is the case, we do not need to consider the array as a high-impedance electromagnetic environment and we only worry about remaining within the small-phase approximation for any SQUID of the array, which amounts to the standard condition $I_b/I_c<<1$, where $I_b$ is the external current bias \cite{simoenthesis}.  We keep this in mind when choosing shape functions.
\begin{figure}
\includegraphics[width=0.5\textwidth]{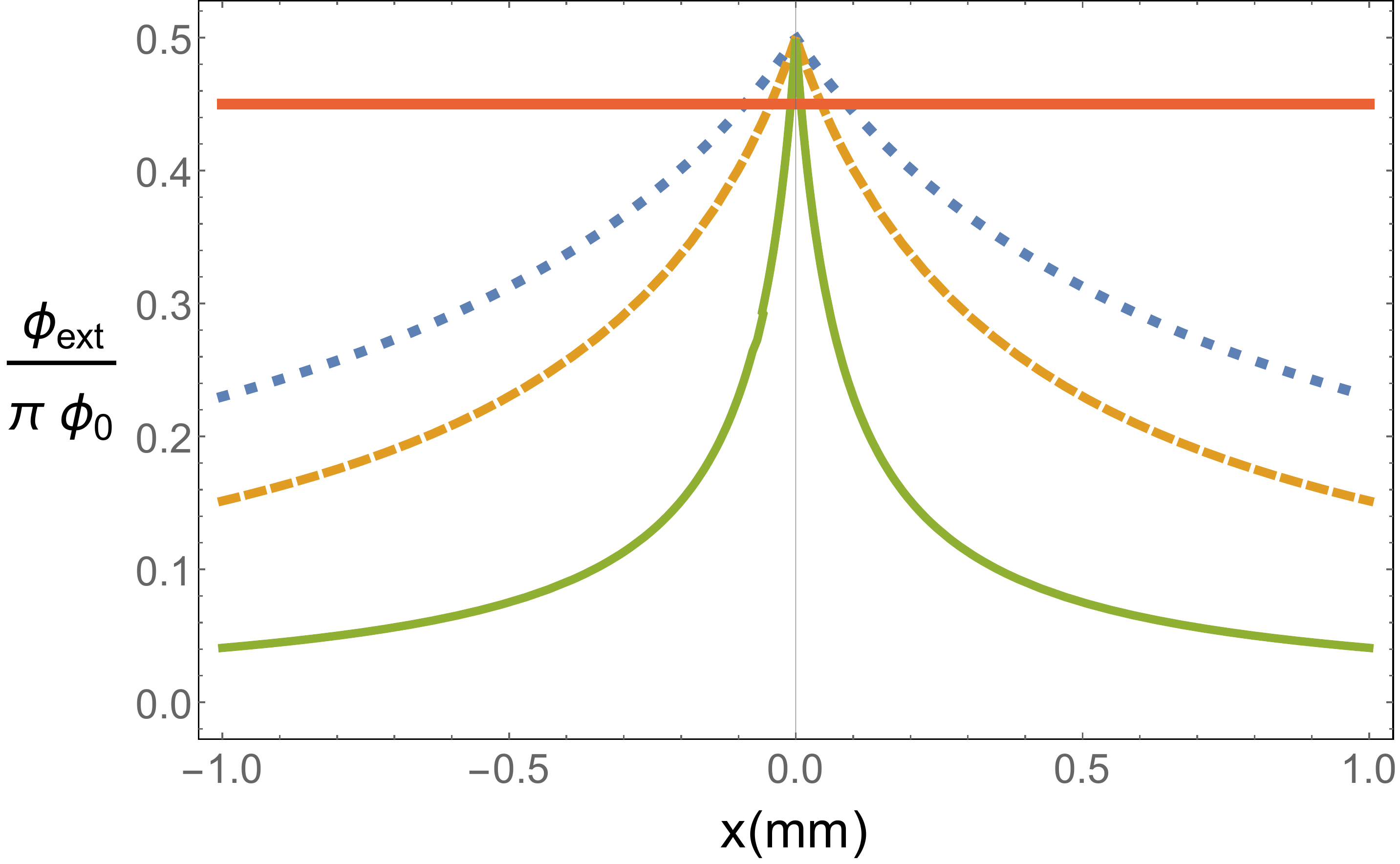}
\caption{(Color online) Flux bias $\phi_{ext}/(\pi\,\phi_0)$ vs. $x (\operatorname{mm})$ as given by Eq. (\ref{eq:externalfluxl}) for different wormhole throat radius $b_0= 1\,\operatorname{mm}$, (blue, dotted), $b_0=0.5\,\operatorname{mm}$ (yellow, dashed) and $b_0=0.1\,\operatorname{mm} $ (green, solid). The value $\phi_{ext}=0.45\pi\phi_0$ (red,solid) is plotted as a reference of critical threshold. The array region above the critical threshold is proportional to the simulated wormhole's throat radius}
\label{fig2}
\end{figure}
\begin{figure}
\includegraphics[width=0.5\textwidth]{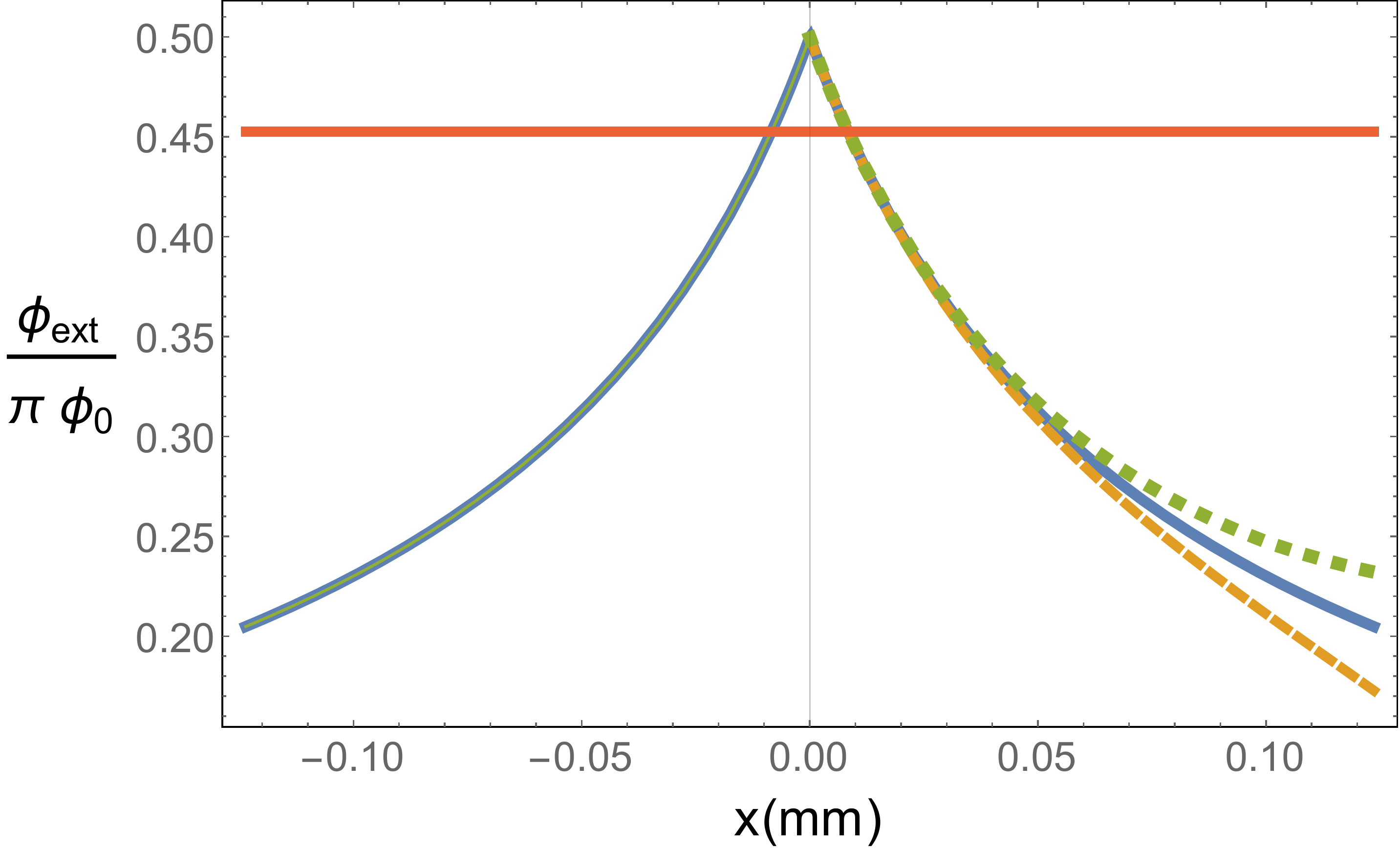}
\caption{(Color online) Flux bias $\phi_{ext}/(\pi\,\phi_0)$ vs. $x (\operatorname{mm})$ as given by Eq. (\ref{fluxtm}) for a wormhole throat radius $b_0= 0.1\,\operatorname{mm}$, $l_0=0.2\,\operatorname{mm}$ and different values of the acceleration $g=0$ (blue, solid), $g=c^2/(20\,l_0)$\, (yellow, dashed) and $g=-c^2/(20\,l_0)$ (green, dotted). The value $\phi_{ext}=0.45\pi\phi_0$ (red,solid) is plotted as a reference of critical threshold. Outside the plotted region, the flux bias should be equal to the corresponding plot in Fig. (\ref{fig2}).}
\label{fig3}
\end{figure}

In order to understand how this 1D array is able to connect two otherwise distant physical regions, we can envision a 2D array consisting of several 1D superconducting transmission lines. For instance, we can place two horizontal standard transmission lines separated by a certain distance $d$, large enough to safely neglect any crosstalk between the lines. In this way, there is of course no way for the microwave field to travel from one transmission line to the other. However, if we place the \textit{wormhole line} vertically interrupting both horizontal lines at some point $x_0$, the 1D microwave fields in the horizontal lines would feel a vertical acceleration when passing by $x_0$ and therefore they would be able to go through the vertical line, thus accessing the other horizontal one.

Of particular interest is the following family of wormholes \cite{ellis,morristhorne, geodesics,taylor}: 
\begin{equation}\label{eq:example}
b(r)=\frac{b_0^2}{r}, 
\end{equation}
for which the proper radial distance to the wormhole throat is simply $l^2(r)=r^2-b_0^2$. Neither the non-monotonic $r$ nor the proper distance $l$ are suitable coordinates to identify them with a position coordinate along the transmission line in the laboratory. To this end, we define a coordinate $x$ such that:
\begin{equation}
|x|=r-b_0,\quad x\in(-\infty,\infty).\label{eq:xcoord}
\end{equation} 
Clearly, $x$ possesses similar features as $l$, since $x=0$ at the wormhole's throat $r=b_0$ and acquires different sign at both sides of the throat. Unlike $l$, it has the advantage that the spacetime metric does not change when transforming coordinates from $r$ to $x$. Notice that $l^2=|x|(|x|+2 b_0)$.
Thus, using Eq. (\ref{eq:externalflux}) and expressing it as a function of $x$, we find:
\begin{equation}\label{eq:externalfluxl}
\phi_{ext} (x)=\frac{\phi_0}{\pi}\operatorname{arccos} (1-\frac{b_0^2}{(|x|+b_0)^2}).
\end{equation}
In Figure \ref{fig2}, we plot Eq. (\ref{eq:externalfluxl}) for several values of the throat radius $b_0$. If we identify $x$ with the position coordinate along the array e. g. we set the wormhole throat $x=0$ in the center of the array, we find that for $b_0=0.1\,\operatorname{mm}$ the flux is below the critical threshold everywhere but in a small region of around $0.02\,\operatorname{mm}$. This could be consistent with the idea of having only one SQUID above the critical value. If the separation $d$ among the SQUIDs is around $0.05 \,\operatorname{mm}$, the array could still be regarded as a continuum for microwave photons up to $200\,\operatorname{GHz}$ ($\lambda=0.5\,\operatorname{mm}>>d$). This frequency cutoff does not represent a stronger limitation than the plasma frequency of the SQUID which is typically smaller than $100\,\operatorname{GHz}$. Notice that while inhomogeneities in the magnetic field bias of the array are usually regarded as problematic and the aim is to minimise them, in our case our goal is to achieve a strongly inhomogeneous field. 

In \cite{morristhorne} it is shown how to turn a traversable wormhole into a {\it time machine}, i.e. a spacetime containing CTC's. The idea is to induce a time shift between the spacetime region at $l>0$ and the one at $l<0$. For instance, one {\it mouth} of the wormhole at $l=l_0$ could be initially at rest with respect to the other mouth at $l=-l_0$ and then follow a twin-paradox trajectory, accelerating up to relativistic speeds in order to travel to a distant star and coming back to the same place. After the trajectory, there is a time shift between $l<0$ and $l>0$ from the point of view of external observers, however if the throat geometry does not change during the trip -which amounts to enforce that $2\,gl_0/c^2<<1$ where $g$ is the maximum acceleration- time does not experience any shift through the throat. Thus, if after the trip an observer travels from $l<0$ to $l>0$ and then back to $l<0$ she would travel along a CTC, accessing in principle her own past. 

The trip of the wormhole mouth in a traversable wormhole spacetime would be codified in the following metric \cite{morristhorne}:
\begin{eqnarray}
ds^2&=&-c^2\,e^{2\Phi(r)}(1+g(t)\,l\,F(l)\cos\theta)^2dt^2+\frac{1}{1-\frac{b(r)}{r}}\,dr^2 \nonumber\\&+&r^2(d\theta^2+\sin^2\theta d\phi^2),\label{eq:metrictm}
\end{eqnarray}
where $g(t)$ is the acceleration and $F(l)$ is a form factor function of the radial distance, vanishing at $l<0$ and rising smoothly up to 1 in the travelling mouth. Again, we consider $\Phi(r)=0$, restrict ourselves to 1D and pull out a conformal factor in order to get the shape of the external flux:
\begin{equation}
\phi_{ext} (r,t)=\frac{\phi_0}{\pi}\operatorname{arccos} (1-\frac{b(r)}{r})(1+g(t)\, l \,F(l))^2,\label{fluxtm}
\end{equation}
where we have further assumed $\theta=0$ for the sake of simplicity. In Fig. (\ref{fig3}) we choose again $b(r)=b_0^2/r$ and also $F(l)= l/l_0$ for $0<l \leq l_0$ and $F(l)=0$ otherwise. We plot the form of the flux in the region between $-x_0$ and $x_0$ for $g=0$, $g=c^2/(20\,l_0)$ and $g=-c^2/(20\,l_0)$, which would characterise the different stages of the accelerated mouth trajectory. Outside this region the flux should be the same as in the case of no acceleration. Notice that if $c=10^8\,\operatorname{m/s}$ and $l_0=0.2\,\operatorname{mm}$, then $g=2.5\cdot10^{18}\,\operatorname{m/s^2}$. The mentioned value of $l_0$ would imply that we only need to adjust the flux of a few SQUIDs, perhaps only 1 if we look at Fig. (\ref{fig3}). For simplicity, we are assuming that the simulated acceleration is instantaneous, meaning that the magnetic flux switches instantaneously among the different curves in Fig. (\ref{fig3}). An abrupt change of the magnetic flux might generate unwanted dynamics, so it would be desirable to include a switching function for the transitions. 

After performing the series of modifications of the flux corresponding to a simulated full twin paradox trajectory for a wormhole mouth, the effective spacetime region between $x_0$ and $-x_0$ contains effective CTCs, which in our case could be probed by sending microwave photons back and forth along this region -e.g. by means of a mirror interrupting the transmission line. If $t$ and $T$ are the times measured by observers at $x_0$ and $-x_0$ respectively, the time shift between them will be given by $T/t=\gamma$ where $\gamma$ is the standard relativistic factor $\gamma=1/\sqrt{(1-v^2/c^2)}$. The relative velocity $v$ in our case will be determined by the acceleration $g$ and the duration of the acceleration as seen by the inertial observer $T_a$, $v=g\,T_a/\sqrt{(1+g^2T_a^2/c^2)}$. Thus, finally: $\gamma=\sqrt{(1+g^2T_a^2/c^2)}$. For the acceleration considered above,  $\gamma\approx25$ after $T_a=1\,\operatorname{ns}$  of acceleration. If the total trajectory lasts $T\,\operatorname{ns}$ as seen by an observer at $-x_0$ this means that the elapsed time for observers at $x_0$ is $T/25\,\operatorname{ns}$, generating a time shift of $24/25 T\,\operatorname{ns}$. Thus photons that have lived in the left side of the transmission line during the acceleration have now the opportunity of travelling back in time $24/25 T\,\operatorname{ns}$ by going first to the right side of the transmission line and back again to the left side. For $T$ of a few $ns$ this time is much larger than the time needed to traverse the  wormhole $t=\int^{x_0}_{-x0}dx/c(x)\simeq 0.04\operatorname{ns}$. The assumption that a photon is travelling along the left part of the transmission line only during a few $\operatorname{ns}$ would imply a transmission line length around $10 \operatorname{cm}$ and thus a number of several thousand SQUIDs.  Transmission lines of more than $2\operatorname{m}$ \cite{gianttlr} and arrays of more than 500 SQUDs \cite{apl} have already been achieved in the laboratory. These numbers have been derived within the assumption that the accelerations are instantaneous. Including realistic switching functions to smooth unwanted dynamics might push them even higher. Notice that in our case there is no \textit{real} acceleration, however the simulated acceleration would give rise to different phase shifts \cite{twinpa} in the microwave field at both sides of the throat. Thus, in our language travelling back in time means acquiring a particular phase shift.

\begin{figure}
\includegraphics[width=0.5\textwidth]{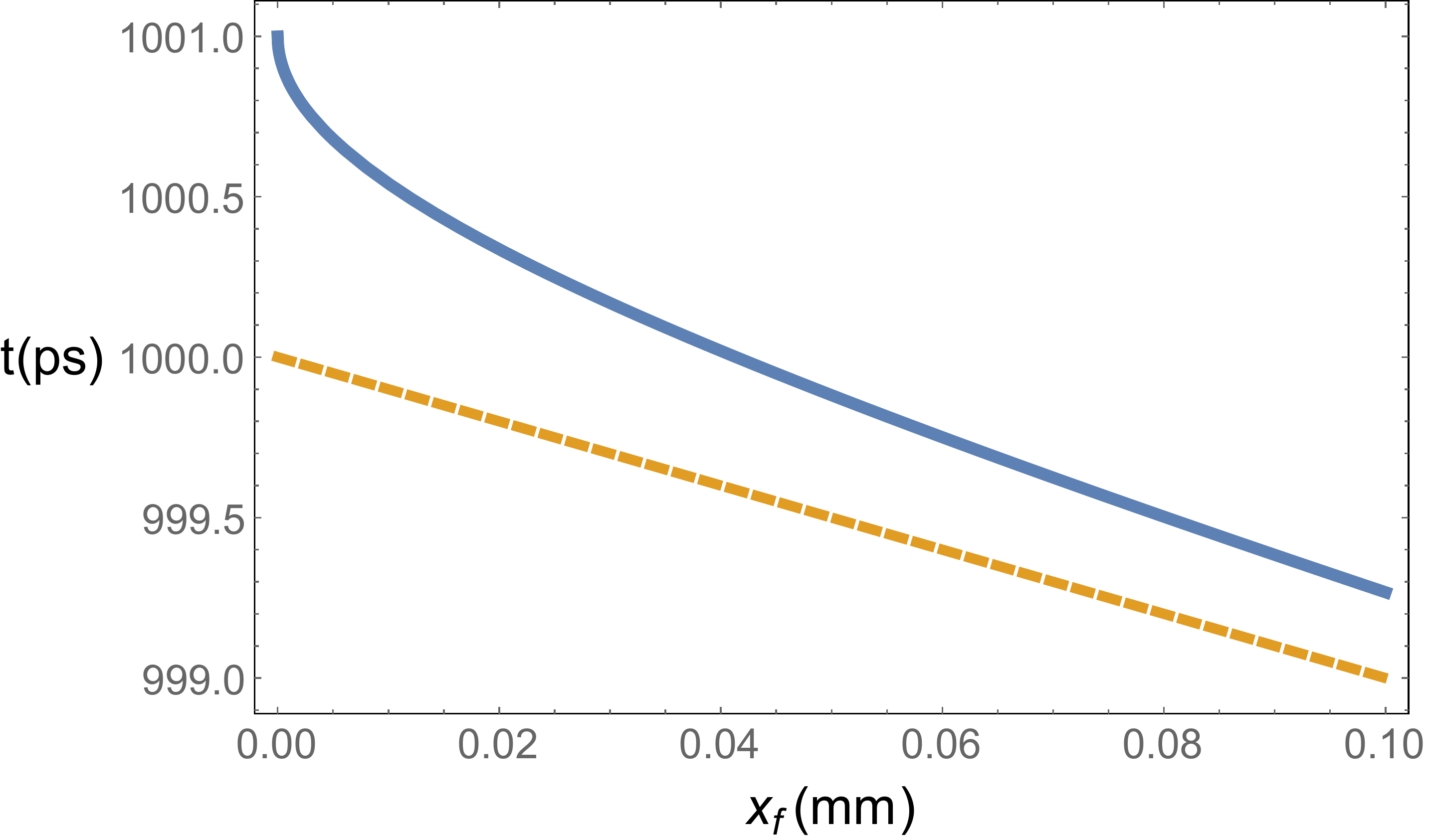}
\caption{(Color online) Elapsed time $t=\int^{x_f}_{x_i}dx/c(x)$ (blue, solid) vs   the final position $x_f (\operatorname{mm})$ for initial position $x_i=10\,\operatorname{cm}$ and throat radius $b_0=0.1\,\operatorname{mm}$ and compared to the corresponding time in the absence of a wormhole $(x_i-x_f)/c$ (yellow, dashed).In both cases $c=10^8\,\operatorname{m/s}$.}
\label{fig4}
\end{figure}

Letting alone the time-machine spacetime, a straightforward way of probing the effective wormhole geometry given by Eq. (\ref{eq:externalfluxl}) would be to measure the time that light takes to travel along the transmission line, which should be slightly delayed with respect to the flat spacetime case. In Fig. (\ref{fig4}) we see that this delay is as high as  $1\,\operatorname{ps}$ after travelling from $x_0=10\,\operatorname{cm}$ to $x_0=0$.

It is interesting to reflect upon the implications of the above results. In \cite{hawking}, Hawking posed the chronology protection conjecture, according to which quantum effects would prevent the formation of CTCs, triggering a fascinating open debate on the subject (see \cite{review} and references therein). Of course, here we do not have a real curved spacetime so Hawking's argument -which is based on divergencies of the quantum propagator of the gravitational energy-momentum tensor- does not apply. However, in our case microwave photons would follow equations of motion that are indistinguishable from the ones of a 1D reduction of the spacetime in Eq. (\ref{eq:metrictm}). Therefore, it is natural to ask: are CTCs forbidden in this cm-size 1D effective spacetime? As a matter of fact, in \cite{reece2015} Reece Boston tried to build up an optical metamaterial containing CTCs -more precisely, closed null geodesics- and finally found that it was actually impossible since the physical parameters of such a metamaterial would be unphysical. In our case, this does not seem to be the case, since there is not anything unphysical in the parameters of the external flux bias, although there are of course important technical challenges. However, we have acknowledged the role of the finite impedance array, which would generate quantum fluctuations of the phase. We have tried to mimimise the impedance so it does not prevent the building-up of a wormhole spacetime. Nevertheless, it is suggestive to think of this effect as an analogue of Hawking's chronology-protection mechanism.  It might be that these quantum fluctuations would prevent us from building an effective wormhole spacetime or from trying to turn it into a time machine. Thus, our quantum simulator could shed light on the operating principles of a chronology protection mechanism. 

If the quantum fluctuations are not strong enough to prevent microwave photons from travelling along effective closed geodesics, this would pave the way to the possibility of using this effect for quantum computing \cite{openctcs} with continuous variables. Moreover, by coupling superconducting qubits to the transmission line we could analyse the entanglement properties of a pair of qubits in the presence of a wormhole background, which would shed light on the conjectured analogy between entanglement and wormholes \cite{maldacena} -in this case, non-traversable ones.

In summary, we provide a recipe to build up an analog quantum simulator of a traversable wormhole 1D background by means of a suitable strongly inhomogeneous external magnetic flux bias along a dc-SQUID array. Furthermore, we show as well how to transform this spacetime in order to allow, in principle, the existence of CTCs. The construction is limited by the quantum fluctuations of the superconducting phase triggered by the array impedance, which might be considered as an analogue of a chronology protection mechanism trying to preserve causality in this tiny 1D effective spacetime for microwave photons.
\section*{Acknowledgements} I am indebted to Borja Peropadre for helpful discussions and comments. Special thanks are given to  Kip Thorne's book \textit{Black holes and time warps: Einstein's outrageous legacy} and Christopher Nolan's movie \textit{Interstellar}. Financial support by Fundaci{\' o}n General CSIC (Programa ComFuturo) is acknowledged.

\end{document}